\documentclass[a4paper]{revtex4}
\begin{document}
\title{State determination: an iterative algorithm.}
\author{Dardo M. Goyeneche and Alberto C. de la Torre }
\email{delatorre@mdp.edu.ar}
 \affiliation{Departamento de F\'{\i}sica,
 Universidad Nacional de Mar del Plata\\
 Funes 3350, 7600 Mar del Plata, Argentina\\CONICET}
\begin{abstract}
An iterative algorithm for state determination is presented that
uses as physical input the probability distributions for the
eigenvalues of two or more observables in an unknown state
$\Phi$. Starting form an arbitrary state $\Psi_{0}$, a succession
of states $\Psi_{n}$ is obtained that converges to $\Phi$ or to a
Pauli partner. This algorithm for state reconstruction is
efficient and robust as is seen in the numerical tests presented
and is a useful tool not only for state determination but also
for the study of Pauli partners. Its main ingredient is the
Physical Imposition Operator that changes any state to have the
same physical properties, with respect to an observable, of
another state.
\end{abstract}
\maketitle
\section{INTRODUCTION}
At an early stage in the development of quantum mechanics, W.
Pauli \cite{paul} raised the question whether the knowledge of
the probability density functions for the position and momentum
of a particle were sufficient on order to determine its state.
That is, can we determine a unique $\psi(x)$ if we are given
$\rho(x)=|\psi(x)|^{2}$ and $\pi(p)=|\phi(p)|^{2}$, where
$\phi(p)$ is the Fourier transform of $\psi(x)$? Since position
and momentum are the unique independent observables of the
system, it was, erroneously, guessed that this Pauli problem
could have an affirmative answer. This was erroneous because
there may be different quantum correlations between position and
momentum that are not reflected in the distributions of position
and momentum individually. Indeed, many examples of Pauli
partners, that is, different states $\psi_{1}\neq\psi_{2}$ with
identical probability distributions $\rho$ and $\pi$, where
found. A review of theses issues, with references to the original
papers, and the treatment of the problem of state reconstruction
for finite and infinite dimension of the Hilbert space, can be
found in refs. \cite{wei1,wei2}. The general problem of the
determination of a quantum state from laboratory measurements
turned out to be a difficult one. In this work the ``laboratory
measurement'' means the complete measurement of an observable,
that is, the determination of the probability distribution
$\varrho(a_{k})$ of the eigenvalues $a_{k}$ of the operator $A$
associated with the observable. Given a state $\Phi$, the
probability distribution (assuming non-degeneracy) is given by
$\varrho(a_{k})=|\langle\varphi_{k},\Phi\rangle|^{2}$ where
$\varphi_{k}$ are the eigenvectors of the operator. The state is
not directly observable; what can be measured, are the
probability distributions  of the eigenvalues of the observables
and we want to be able to determine the state $\Phi$ of the
system using these distributions. Besides the academic interest
of quantum state reconstruction based on measurements of
probability distributions, the issue has gained actuality in the
last decade in the possible practical applications of quantum
information theory \cite{qinf}.

In order to state clearly the problem, let us consider a system
described in an $N$ dimensional Hilbert space. The determination
of the state requires the determination of $2N-2$ real numbers
and a complete measurement of an observable provides $N-1$
equations. With the measurement of two observables (like position
and momentum in the Pauli problem) we have the same number of
equations as unknowns. However the equations available are not
linear and the system of equations will not have, in general, a
unique solution. In many practical cases, a minimal additional
information (like the sign of an expectation value) is sufficient
to determine the state. In this work we will not search the
minimal extra information required, but instead, we will add a
complete measurement of a third observable. One may think that
this massive addition of information will make the system
over-determined and that with three complete measurements we
should always be able to find a unique state. This is wrong;
there are pathological cases where the complete measurement of
$N$ observables, that is $N(N-1)$ equations, is \emph{not
sufficient} for the determination of a \emph{unique} set of
$2(N-1)$ numbers! In the other extreme, if the state happens to
be equal to one of the eigenvectors of the observable measured,
then, of course, just one complete measurement is sufficient to
fix the state. From these two cases we conclude that the choice
of the observables to be measured is crucial for the
determination of the state; an observable with a probability
distribution peaked provides much more information than an
observable with uniform distribution. A pair of observables may
provide redundant information and we expect that it is convenient
to use observables as different as possible; this happens when
their eigenvectors build two unbiased bases as is the case, for
example, with position and momentum (two bases $\{\varphi_{k}\}$
and $\{\phi_{r}\}$ are unbiased when
$|\langle\varphi_{k},\phi_{r}\rangle|=1/\sqrt{N}\ \forall k,r$,
that is, every element of one basis has equal ``projection'' on
all elements of the other basis). For this reason, unbiased bases
have been intensively studied in the problem of state
determination and also in quantum information theory
\cite{unbiaBas1,unbiaBas2,unbiaBas3}. The number of mutually
unbiased bases that one can define in an $N$ dimensional Hilbert
space is not known in general although it can be proved that if
$N$ is equal to a power of a prime number, then there are $N+1$
unbiased bases. \emph{Unbiased observables}, those represented by
operators whose eigenvectors build unbiased bases, provide
independent information; there are however pathological cases
where the measurement of several unbiased observables is useless
to determine a unique state: assume for instance that the state
belongs to a basis that is unbiased to several other mutually
unbiased bases associated with the measured observables. In this
case all the probability distributions are uniform and the state
can not be uniquely determined because there are at least $N$
different states (corresponding to the $N$ elements of the basis
to which the state belongs) all generating uniform distributions
for the observables. If $N$ is a power of a prime number we could
have up to $N$ observables with uniform distributions for $N$
different states. This is the pathological case mentioned before:
if there are $N+1$ mutually unbiased bases and we have $M$
unbiased observables with uniform distributions then we have
$N(N+1-M)$ Pauli partners, that is, different states having the
same distributions.

If we make complete measurements of two or more observables we
should be able to determine the state but it will not always be
unique because there may be several different states having the
same distributions for the measured observables. If we measure
three observables, the mathematical problem would be to solve a
set of $3N-3$ nonlinear equations to determine $2N-2$ numbers.
One could blindly apply some numerical method to find the
solution. Instead of this, we present in this work an iterative
method that is physically appealing because it involves the
imposition of physical data to Hilbert space elements that are
approaching the solution. Another advantage of this algorithm is
that it does not involves the solution of a system of equations
and therefore when we change the number of observables measured
or the dimension of the Hilbert space, we only have to make a
trivial change in the algorithm. We will test the algorithm
numerically by assuming an arbitrary state and two, three or four
arbitrary observables, with them we generate the data
corresponding to the distributions of the observables in the
chosen state, and then we run the algorithm and we see how
efficiently it returns the chosen state.
\section{A METRIC FOR STATES}
In order to study the convergence of an iterative algorithm for
the determination of a state, we will need a concept of
\emph{distance} that can tell us how close we are from the wanted
solution. This criteria of approach can be applied in the space
of states or in the space of probability distributions. In the
first case we want to know how close a particular state is from
the state searched, that is, we need a \emph{metric in the space
of states}. In the other case a particular state generates
probability distributions for some observables and we want to
know how close these distribution are from the corresponding
distributions generated by the state searched. In this second
case we need a \emph{metric in the space of distributions}. The
relation between these two distances in two different spaces has
been studied for several choices of distances \cite{plast}.
However the application of some of theses ``distances'' that do
not satisfy the mathematical requirements of a ``metric''
(positivity, symmetry and triangular inequality) in an iterative
algorithm is questionable. In this work we use a metric in the
Hilbert space of states in order to study the convergence of the
algorithm but we also compare the final probability distributions
with the corresponding distributions used as physical input for
the algorithm because, as was explained before, there are cases
of different states generating the same probability
distributions. The usual Hilbert space metric induced by the
norm, itself induced by the internal product,
\begin{equation}\label{metrind}
\delta(\Psi,\Phi)=\|\Psi-\Phi\|=\sqrt{\langle\Psi-\Phi,\Psi-\Phi\rangle}
\end{equation}
is not an appropriate metric for states because the states are
nor represented by Hilbert space elements but by \emph{rays},
that are sets of Hilbert space elements with an arbitrary phase.
That is, a state is given by
\begin{equation}\label{sta}
   R_{\Psi}=\{e^{i\alpha}\Psi\ |\ \forall \alpha\ ,\ \|\Psi\|=1\}
\end{equation}
The Hilbert space element $\Psi$ is a \emph{representant} of the
ray and it is common practice in quantum mechanics to say that
the state is given by $\Psi$. However when we deal with distance
between states we can not take the induced metric mentioned
before, because this metric for two Hilbert space elements,
$e^{i\alpha}\phi$ and $e^{i\beta}\phi$, belonging to the
\emph{same} ray, that is, belonging to the same state, does not
vanishes. A correct concept of distance between states is given
by the distance between sets
\begin{equation}\label{disray}
 d(R_{\Psi},R_{\Phi})=\min_{\alpha,\beta}\delta(e^{i\alpha}\Psi,e^{i\beta}\Phi)\
 .
\end{equation}
 The minimization can be performed in general and we obtain
\begin{equation}\label{disray1}
 d(R_{\Psi},R_{\Phi})=\sqrt{2}\sqrt{1-|\langle\Psi,\Phi\rangle|}\ .
\end{equation}
We compare this result with
$\delta(\Psi,\Phi)=\sqrt{2}\sqrt{1-\Re\langle\Psi,\Phi\rangle}$
and we conclude that $d(R_{\Psi},R_{\Phi})\leq\delta(\Psi,\Phi)$
and therefore every sequence converging in the induced metric is
also convergent in the ray metric used here.

In order to have a rigourous concept of convergence we must check
that the distance between states given above in
Eqs.(\ref{disray},\ref{disray1}) is really a metric (in general,
for arbitrary sets, the distance between sets is not always a
metric since one can easily find examples that violate the
triangle inequality). The requirement of symmetry and positivity
are trivially satisfied but to prove that this distance satisfies
the triangle inequality is not trivial. However we can be sure
that the distance between states is a metric because, in this
particular case where the sets are rays, the distance between
rays has the same value as the Hausdorff distance and one can
prove that the Hausdorff distance is a metric \cite{haus}. The
Hausdorff distance between two sets $X$ and $Y$ is defined by
\begin{equation}\label{hausdm}
  d^{H}(X,Y)=\max\left\{\sup_{x\in X}d(x,Y)\ ,\ \sup_{y\in
  Y}d(y,X)\right\}\ .
\end{equation}
As a final comment in this section, notice that the square root
in Eq.(\ref{disray1}) is a nuisance but it can not be avoided
because expressions like $1-|\langle\Psi,\Phi\rangle|$ or
$1-|\langle\Psi,\Phi\rangle|^{2}$ are \emph{not} metrics. In
order to simplify the notation, in what follows, we will denote
the distance between rays $d(R_{\Psi},R_{\Phi})$ simply by
$d(\Psi,\Phi)$.

\section{THE PHYSICAL IMPOSITION OPERATOR} A state, or a Hilbert space element,
contains encoded information about all the observables of the
system. Given a state $\Phi$, the probability distribution for an
observable $A$ is given by
$\varrho(a_{k})=|\langle\varphi_{k},\Phi\rangle|^{2}$ where
$\varphi_{k}$ are the eigenvectors of the  operator $A$
associated with the observable corresponding to the eigenvalue
$a_{k}$. Given any state $\Psi$, we can impose to this state the
same distribution that the observable $A$ has in the state $\Phi$
by means of an operator, the \emph{Physical Imposition Operator},
$T_{A\Phi}$ that involves the expansion of $\Psi$ in the basis
$\{\varphi_{k}\}$ of $A$ and a change in the modulus of the
expansion coefficients. That is
\begin{equation}\label{TAfi}
    T_{A\Phi}\Psi = \sum_{k}|\langle\varphi_{k},\Phi\rangle|
    \frac{\langle\varphi_{k},\Psi\rangle}{|\langle\varphi_{k},\Psi\rangle|}
    \ \varphi_{k}\ .
\end{equation}
If $\langle\varphi_{k},\Psi\rangle=0$ we assume zero phase, that
is
$\langle\varphi_{k},\Psi\rangle/|\langle\varphi_{k},\Psi\rangle|
= 1$. The moduli of the expansion coefficients are changed in
order to impose the distribution of the observable $A$ in the
state $\Phi$ but the phases are retained and therefore some
information of the original state $\Psi$ is kept in the phases.
Although the numerical treatment of this operator is
straightforward, its mathematical features are not simple. The
operator is idempotent $T^{2}=T$, it has no inverse and it is not
linear but $T(c\Psi)=(c/|c|)T(\Psi)$. Furthermore the operator is
bounded because $\|T\Psi\|=\|\Phi\|=1$. The fix points of this
nonlinear application is the set of states that have the same
distribution for the observable $A$ as the state $\Phi$.

We will use this operator in order to develop an iterative
algorithm for the determination of a state $\Phi$ using as
physical input the distribution of several observables in this
state. It is therefore interesting to study whether this
operator, applied to an arbitrary Hilbert space element $\Psi$,
brings us closer to the state $\Phi$ or not. For this we can
compare the distance $d(\Psi,\Phi)$ with the distance
$d(T_{A\Phi}\Psi,\Phi)$ for some given observable $A$ and some
state $\Phi$. Let us then define an observable $A$ by choosing
its eigenvectors $\{\varphi_{k}\}$ (a basis) in a three
dimensional Hilbert space, $N=3$, and in this space let us take
an arbitrary state $\Phi$. Now we consider a large number (8000)
of randomly chosen states $\Psi$ and draw a scatter plot of the
distances of this state to $\Phi$ before and after applying the
imposition operator $T_{A,\Phi}$. In Figure 1 we see that there
are more points below the diagonal, showing cases where the
imposition operator brings us closer to the state but there are
also many cases where the operator take us farther away from the
searched state. We will later see that this has the consequence
that the iterative algorithm will not converge for every starting
point.

The imposition operator will shift the state $\Psi$ some distance
$d(T_{A\Phi}\Psi,\Psi)$ that is smaller than the total distance
to the state $d(\Psi,\Phi)$. That is, there is no ``overshoot''
that could undermine the convergence of the iterative algorithm.
In order to prove this, consider the internal product
\begin{eqnarray}
 \nonumber
  \left\langle T_{A\Phi}\Psi,\Psi\right\rangle&=& \left\langle
  \sum_{k}|\langle\varphi_{k},\Phi\rangle|
    \frac{\langle\varphi_{k},\Psi\rangle}{|\langle\varphi_{k},\Psi\rangle|}
    \ \varphi_{k}\ ,\ \sum_{r}\langle\varphi_{r},\Psi\rangle
    \ \varphi_{r}\right\rangle\\ \nonumber
   &=&\sum_{k}|\langle\varphi_{k},\Phi\rangle|\
    |\langle\varphi_{k},\Psi\rangle|\ =
    \sum_{k}|\langle\Phi,\varphi_{k}\rangle\langle\varphi_{k},\Psi\rangle|\ \\
  &\geq&|\sum_{k}\langle\Phi,\varphi_{k}\rangle\langle\varphi_{k},\Psi\rangle|\
  = |\langle\Phi,\Psi\rangle|\ .
\end{eqnarray}
Now, using this inequality in the definition of distance in
Eq.(\ref{disray1}) we get\begin{equation}\label{dist1}
  d(T_{A\Phi}\Psi,\Psi)\leq d(\Psi,\Phi)\ .
\end{equation}
We can notice in Figure 1 that there is a bound for the distance
$d(T_{A\Phi}\Psi,\Phi)$ at some value smaller than the absolute
bound for the distance $\sqrt{2}$. We will see that this bound
appears when the state $\Phi$ is chosen close to one of the
eigenvectors of $A$. From the definition of the distance and of
the imposition operator it follows easily that the distance
 of $T_{A\Phi}\Psi$ to any element of
$\{\varphi_{k}\}$ is the same as the distance of $\Phi$ to the
same element. That is,
\begin{equation}\label{dist2}
  d(T_{A\Phi}\Psi,\varphi_{k}) =d(\Phi,\varphi_{k})\  \forall k\
  ,
\end{equation}
so that $T_{A\Phi}\Psi$ is something like a ``mirror image'' of
$\Phi$ reflected on $\{\varphi_{k}\}$. We can now use this in
order to derive the bound mentioned. Consider the triangle
inequality $d(T_{A\Phi}\Psi,\Phi)\leq
d(T_{A\Phi}\Psi,\varphi_{k})+d(\Phi,\varphi_{k})$. Using
Eq.(\ref{dist2}), we get $d(T_{A\Phi}\Psi,\Phi)\leq 2
d(\Phi,\varphi_{k})$. Now we specialize this inequality for the
value of $k$ that minimizes the right hand side, that is, the
value of $k$ that maximizes $|\langle\varphi_{k},\Phi\rangle|$ or
equivalently $\sqrt{\rho(a_{k})}$. Then we have
\begin{equation}\label{dist3}
d(T_{A\Phi}\Psi,\Phi)\leq 2 \min_{k}d(\Phi,\varphi_{k})
=2\sqrt{2}\sqrt{1-\max_{k}\sqrt{\rho(a_{k})}}\ .
\end{equation}

If the state $\Phi$ is close enough to one of the eigenvectors of
$A$, the corresponding maximum value of the distribution can be
larger than $9/16$ and the bound derived is smaller than the
absolute bound $\sqrt{2}$. With increasing dimension $N$ of the
Hilbert space, the probability that a randomly chosen state is
close to one of the basis elements decreases.

The physical imposition operator modifies the moduli of the
expansion coefficients but leaves the phases unchanged. The
reason for choosing this definition is that the moduli of the
coefficients are measured in an experimental determination of the
probability distribution of the eigenvalues of an observable and
therefore this operator provides a way to impose physical
properties to a state. It is unfortunate that the phases of the
expansion coefficients are not directly accessible in an
experiment because we could use the knowledge of the phases in a
much more efficient algorithm. In a sense that will become clear
later, the phases have more information about the state than the
moduli. In order to clarify this let us define a \emph{Phase
Imposition Operator} $P_{A\Phi}$ that leaves the moduli of the
expansion coefficients unchanged but imposes the phases of the
state $\Phi$. That is
\begin{equation}\label{RAfi}
    P_{A\Phi}\Psi = \sum_{k}|\langle\varphi_{k},\Psi\rangle|
    \frac{\langle\varphi_{k},\Phi\rangle}{|\langle\varphi_{k},\Phi\rangle|}
    \ \varphi_{k}\ .
\end{equation}
The same as was done before, we study how efficiently this
operator approaches to the state $\Phi$. In Figure 2  we see the
corresponding scatter plot for the same operator and states of
those in Fig.1, that shows that in \emph{all} cases the
application of this operator brings us closer to the wanted
state. One can indeed prove that $d(P_{A\Phi}\Psi,\Phi)\leq
d(\Psi,\Phi)$ considering the internal product
\begin{eqnarray}
 \nonumber
  \left\langle P_{A\Phi}\Psi,\Phi\right\rangle&=& \left\langle
  \sum_{k}|\langle\varphi_{k},\Psi\rangle|
    \frac{\langle\varphi_{k},\Phi\rangle}{|\langle\varphi_{k},\Phi\rangle|}
    \ \varphi_{k} ,\ \sum_{r}\langle\varphi_{r},\Phi\rangle
    \ \varphi_{r}\right\rangle\\ \nonumber
   &=&\sum_{k}|\langle\varphi_{k},\Psi\rangle|\
    |\langle\varphi_{k},\Phi\rangle|\ =
    \sum_{k}|\langle\Psi,\varphi_{k}\rangle\langle\varphi_{k},\Phi\rangle|\ \\
  &\geq&|\sum_{k}\langle\Psi,\varphi_{k}\rangle\langle\varphi_{k},\Phi\rangle|\
  = |\langle\Psi,\Phi\rangle|\ .
\end{eqnarray}
Using this inequality in the definition of distance in
Eq.(\ref{disray1}) we get the inequality above. As said before,
if we had physical information about the phases of the expansion
coefficients, we could devise a very efficient algorithm.
Unfortunately we don't have experimental access to the phases and
this, in principle interesting, operator will no be further
studied here.
\section{THE ALGORITHM FOR STATE DETERMINATION}
In this section we will investigate an algorithm for state
determination that uses as physical input the knowledge provided
by the complete measurement of several observables. These
measurements provide the probability distributions for the
eigenvalues in the unknown state $\Phi$. In other words, we
assume that we know the physical imposition operators
$T_{A\Phi},T_{B\Phi},T_{C\Phi}\dots$ for several observables. The
algorithm basically consists in the iterative application of the
physical imposition operators to some arbitrary initial state
$\Psi_{0}$ randomly chosen. Applying the operator $T_{A\Phi}$ to
the initial state $\Psi_{0}$, we will get closer to $\Phi$
(although not always) but a second application of the operator is
useless because $T_{A\Phi}$ is idempotent. Then we use another
operator for a closer approach, say $T_{B\Phi}$, and another one
afterwards, until all physical information is used; then we start
again with $T_{A\Phi}$. That is, we calculate the iterations
$\Psi_{1},\Psi_{2},\Psi_{3}\cdots$ given by $\Psi_{n}=(\dots
T_{B\Phi}T_{A\Phi})^{n}\Psi_{0}$ and the convergence
$\Psi_{n}\rightarrow\Phi$ is checked comparing the physical
input, that is, the distributions associated with the observables
$A,B,C,\dots$, with the corresponding distributions generated in
the state $\Psi_{n}$.

In order to check the efficiency of the algorithm numerically, we
choose a state $\Phi$ at random and with it we generate the
distributions corresponding to some observables that we use as
input in the algorithm. Calculating the distance
$d(\Psi_{n},\Phi)$ we study how efficiently the algorithm returns
the initial state $\Phi$. There are cases where the algorithm
converges to a state $\Phi'$ different from $\Phi$ but having the
same physical distributions, that is, to a Pauli partner of
$\Phi$. An interesting feature of the algorithm is that we can
span the whole Hilbert space by choosing the starting states
$\Psi_{0}$ randomly and the algorithm delivers many, if not all,
Pauli partners. We can not be sure that all Pauli partners are
found because some of them could correspond to a starting state
$\Psi_{0}$ belonging to a set of null measure that will not be
necessarily ``touched'' in a random sampling of the Hilbert
space. This seems to be quite unlikely but it can not be
excluded. In this way, the algorithm presented is not only a
numerical tool for state determination but is also a useful tool
for the theoretical investigation of the appearance of Pauli
partners. An example of this is presented below.

The algorithm is very efficient; however there are some starting
states $\Psi_{0}$ where the algorithm fails to converge. It was
not surprising to find these failures because, as was suggested
in Fig. 1, the physical imposition operator sometimes take us
farther away from the wanted state. We are informed of this
failure because the distributions used as input are not
approached in each iteration. In the case of a failure, we can
simply restart the algorithm with a different initial state
$\Psi_{0}$ or restart with another initial state orthogonal to
the one that failed. In this last case the probability of a
repeated failure is much reduced and therefore it is a convenient
choice. The appearance of a failure depends strongly on the
choice of observables used to determine the state. If we use
three unbiased observables we very rarely found a failure, in
less than 1\% of the cases,  but if we use three random
observables (see below) 40\% of the randomly chosen starting
states $\Psi_{0}$ fail but only 10\% of these fail again if we
restart with an orthogonal state. In the case of four angular
momentum observables in four arbitrary directions we had to
restart the algorithm in some 10\% of the cases. The appearance
of failures also depends on the shape of the distributions: when
one of the distributions is peaked, that is, the maximum value of
the distribution $\rho(a_{k})$ has a large value for some $k$,
the application of the corresponding imposition operator bring us
close to the wanted state as can be seen in Eq.(\ref{dist3}), and
Figure 1 shows that then the algorithm has better convergence and
no failures are found. This has been confirmed in the numerical
tests.

The convergence to the state $\Phi$, or to a Pauli partner, was
tested numerically in several Hilbert space dimensions and for
different choice of observables. These choices were random in
some cases, that is, their associated orthonormal bases are
randomly chosen, and in other cases we used physically relevant
observables like angular momentum or position and momentum.
Position and momentum observables are usually represented by
unbound operators in infinite dimensional Hilbert spaces; however
there are also realizations of these observables in finite
dimensions, for instance in a cyclic lattice, where they are
represented by unbiased operators \cite{findim1,findim2}. In
general the operators $T_{A\Phi},\ T_{B\Phi},\cdots $ do not
commute and the iteration of $(\cdots T_{B\Phi}T_{A\Phi})^{n}$
and $(\cdots T_{A\Phi}T_{B\Phi})^{n}$ are not necessarily equal.
The algorithm was tested with several different choices in the
ordering of the noncommuting physical imposition operators and
also with random ordering  and it turned out that the convergence
of the algorithm is not much affected by the different orderings.
The algorithm is robust under the noncommutativity of the
observables.

The physical imposition operator $T_{A\Phi}$ is idempotent so it
is useless to apply it more than once (successively) in an
attempt to approach the state $\Phi$. Clearly, the complete
measurement of just one observable is not sufficient to determine
the state, except in the trivial case when the state happens to
be equal to one of the eigenvector of the operator. Therefore we
consider the information provided by \emph{two} observables $A$
and $B$ (for two unbiased observables, like $X$ and $P$, this is
precisely the Pauli problem). We studied then the convergence of
$\Psi_{n}=(T_{B\Phi}T_{A\Phi})^{n}\Psi_{0}$ towards $\Phi$ or to
a Pauli partner, for an arbitrary $\Psi_{0}$. In a three
dimensional Hilbert space, $N=3$, we applied the algorithm in
several cases: for $A$ and $B$ random, unbiased (that is, of the
type $X$ and $P$)  and also for angular momentum operators
$J_{x},J_{y}$. As was expected, in all these cases the algorithm
returned several Pauli partners. Choosing the starting state
$\Psi_{0}$ randomly (uniform distributed in the Hilbert space) we
found that all Pauli partners found are accessed with similar
frequency. As was mentioned before, we can not be sure that the
algorithm will deliver \emph{all} partners, however we may be
confident that this may be so because in one particular case,
where we can calculate exactly the number of partners, the
algorithm returns them all. The particular case is the, so
called, pathological case where we have uniform distributions for
$M$ observables that correspond to $N(N+1-M)$ partners. For
several combinations of $N$ and $M$, the algorithm delivered all
partners.

Next we studied the case with \emph{tree} operators providing
physical information to determine the state (also with Hilbert
space dimension $N=3$). We studied then the iteration
$\Psi_{n}=(T_{A\Phi}T_{B\Phi}T_{C\Phi})^{n}\Psi_{0}$. When two of
the observables are unbiased (of the type $X$ and $P$) we always
obtained a unique state, regardless of the choice of the third
operator: either unbiased or of the type $X+P$ (biased to the
first two), or random. This means that the information provided
by two unbiased observables \emph{almost} fixes the state and any
other additional information is sufficient to find a unique
state. However we know that in the, so called, pathological cases
we must find Pauli partners and the algorithm does indeed finds
them. In these pathological cases, the distributions
corresponding to three unbiased operators are all uniform (that
is, the generating state $\Phi$ is unbiased to all three bases).
Spanning the Hilbert space by choosing $\Psi_{0}$ randomly as a
starting state for the algorithm, we converge to all $N(N-2)=3$
Pauli partners with almost equal probability. The pathological
case was also studied with two unbiased observables with uniform
distributions. In this case the algorithm also delivered all
$N(N-1)=6$ Pauli partners with similar probability.

For biassed operators, like angular momentum operators
$J_{x},J_{y},J_{z}$ and also for random $A, B, C$ we sometimes
found Pauli partners showing that, although we have more
equations (six) than unknowns (four), the nonlinearity of the
problem may cause non-unique solutions. The appearance of Pauli
partners in the angular momentum case is consistent with the
result reported by Amiet and Weigert \cite{wei4}. An inspection
of the numerical results for these Pauli partners revealed a
symmetry that could also be proved analytically: given a state
$\phi$ (in the basis of $J_{z}$) with the corresponding
distributions for the observables $J_{x},J_{y},J_{z}$,
\begin{equation}\label{partn}
 \phi=   \left(%
\begin{array}{c}
 a \\
  b \\
 c\\
\end{array}%
\right) \ ,
\end{equation}
(it is always possible to fix $b$ real an nonnegative) when $b>0$
and $(a^{\ast}+c)\neq 0$, if any one of the following conditions
is satisfied:
\begin{eqnarray}
\Re(a)  &=& -\Re(c)\ , \\
  \Im(a) &=& \Im(c)\ , \\
  |a|&=& |c|\ , \\
  \Im(ac) &=& 0\ ,
\end{eqnarray}
then there is a Pauli partner $\phi'$
\begin{equation}\label{partn1}
\phi'= \left(%
\begin{array}{c}
 a' \\
 b\\
 c'\\
\end{array}%
\right)\ ,
\end{equation}
where
\begin{equation}\label{partn2}
   a'= a^{\ast}\frac{(a+c^{\ast})}{(a^{\ast}+c)}\ \ ,\
   c'= c^{\ast}\frac{(a^{\ast}+c)}{(a+c^{\ast})} \ .
\end{equation}
If $b=0$ we can make $a$ real and positive and then $a'=a\ ,\
c'=c^{\ast}$, and finally if $(a^{\ast}+c)= 0 $ then
$c'=-a'^{\ast}$, where $a'$ can take three values: $-a\ ,\
ia^{\ast}\ , \ -ia^{\ast}$. Spanning the Hilbert space with
generator states $\phi$ randomly chosen, in some $1\%$ of the
cases the algorithm returned the state $\phi$ and a Pauli partner
$\phi'$ covering all possibilities mentioned above. Notice that
the ability of the algorithm to detect Pauli partners is due to
the limited precision of the numerical procedure. Among all
possible states $\phi$ of the system, only a few of them have
Pauli partners,  more precisely, the set of states with Pauli
partners has null measure and if we had infinite precision, we
would never find partners by random sampling of the Hilbert
space. Because of the limited precision of the algorithm, all
points in the Hilbert space within a small environment are
equivalent and therefore the sets of points with null measure can
be accessed in a random sampling of the Hilbert space. We have
found indeed that if we become more restrictive with conditions
of convergence we need more tries in order to detect partners.
Usually the limited precision is considered a drawback however in
this case it is an advantage that allows us to detect sets of
null measure.

 With the information provided by the complete
measurement of \emph{four} operators $A, B, C, D$ we iterated
$\Psi_{n}=(T_{A\Phi}T_{B\Phi}T_{C\Phi}T_{D\Phi})^{n}\Psi_{0}$ and
we found unique states, not only when two of them are unbiased
(consistent with the result obtained with three operators), but
also in the case of random operators or angular momentum in
\emph{arbitrary} directions $J_{r},J_{s},J_{t},J_{u}$. Of course,
in this case of excessive physical information we could ignore
one of the observables and determine the state with only three of
them. However not all the Pauli partners found with three
observables will have the correct distribution for the fourth one
and therefore the use of all observables may be needed for a
unique determination of the state. In this case the number of
equations, eight if $N=3$, uniquely determine the four unknowns
in spite of the nonlinearity. Notice that the convergence of the
algorithm in this case is not trivial. It is true that we are
using much more information than what is needed (except for the
pathological cases that can only appear if $N>3$) but we must
consider that we are using this excessive information in an
iterative and approximative algorithm and therefore the
consistency of the data in the final state does not necessarily
cooperates in the iterations. The fact that the over-determined
algorithm converges is a sign of its robustness.

The algorithm converges in a very efficient way, close to
exponential, as we see in Figure 3 where the distance to the
converging state is given as a function of the number of
iterations for the case of three unbiased operators with $N=3$.
This is a typical example showing the exponential convergence
where the distance to the solution is divided by 4.5 in each
iteration. However the speed of convergence, that is, the slope
in the figure, is not always the same and depends on the
operators used and on the generating state $\Phi$. For higher
Hilbert space dimensions we obtained similar behaviour. For three
operators with physical relevance, like angular momentum or
unbiased operators, the distance to the target state was divided
by 2-3 in each iteration in Hilbert spaces with dimensions up to
20. In the fastest case found, the distance was divided by 126 in
each iteration, approaching the solution within $10^{-7}$ after
three iterations. With random operators the approach was not
always so fast and in some unfavourable cases up to 100
iterations were required (this took only a fraction of a second
in an old PC).
\section{CONCLUSION}
In this work we defined the \emph{Physical Imposition Operator}
$T_{A\Phi}$ that imposes to any state $\Psi$ the same
distribution that the eigenvalues of an observable $A$ have in a
state $\Phi$. For this operator we don't need to know the state
$\Phi$ but we just need the probability distribution for the
observable $A$ in  this state, that can be obtained from a
complete measurement. Considering two or more observables, we
applied their corresponding physical imposition operators
iteratively to an arbitrary initial state $\Psi_{0}$ and obtained
a succession of states $\Psi_{n}$ that converge to the unknown
state $\Phi$, or to a Pauli partner having the same distribution
for the observables. Varying the initial state we can find the
Pauli partners but we can not be sure that all of them are
obtained although this is very likely because in the cases where
we can know exactly all the Pauli partners, the algorithm finds
them all and therefore it becomes a useful tool for the
investigation of Pauli partners. This algorithm for state
determination was tested numerically for different sets of
observables and different dimensions of the Hilbert space and it
turned out to be quite an efficient and robust way to determine a
quantum state using complete measurements of several observables.
\section{Acknowledgements}
We would like to thank H. de Pascuale for his help on
mathematical questions. This work received partial support from
``Consejo Nacional de Investigaciones Cient{\'\i}ficas y T{\'e}cnicas''
(CONICET), Argentina. This work, part of the PhD thesis of DMG,
was financed by a scholarship granted by CONICET.

\newpage
\section{FIGURE CAPTIONS}
\noindent FIGURE 1. Scatter plot of the distances $d(\Psi,\Phi)$
and $d(T_{A\Phi}\Psi,\Phi)$ for 8000 random initial states
$\Psi$. Points below the diagonal indicate cases where
$T_{A\Phi}$ brings $\Psi$ closer to $\Phi$.\\ \\ \\
 FIGURE 2. Scatter plot of the distances $d(\Psi,\Phi)$ and $d(P_{A\Phi}\Psi,\Phi)$
for the same operator and states as in Figure 1. Notice that the
Phase Imposition Operator $P_{A\Phi}$ always approaches the state
$\Phi$.\\ \\ \\
 FIGURE 3. Distance from the state
 $\Psi_{n}=(T_{A\Phi}T_{B\Phi}T_{C\Phi})^{n}\Psi_{0}=T^{n}\Psi_{0}$ to the
 state $\Phi$ after $n$ iterations, showing exponential
 convergence of the algorithm for $A,B,C$ unbiased operators
 in a three dimensional Hilbert space.


\begin{thebibliography}{99}
\bibitem{paul}
W. Pauli. ``Quantentheorie'' Handbuch der Physik {\bf 24}(1933).
\bibitem{wei1}
S. Weigert. ``Pauli problem for a spin of arbitrary length: A
simple method to determine its wave function'' Phys Rev. A {\bf
45}, 7688-7696 (1992).
\bibitem{wei2}
S. Weigert. ``How to determine a quantum state by measurements:
The Pauli problem for a particle with arbitrary potential'' Phys
Rev. A {\bf 53}, 2078-2083 (1996).
\bibitem{qinf}M. Keyl,
``Fundamentals of quantum information theory'' Phys. Rep. A
\textbf{369}, 431-548 (2002).
\bibitem{unbiaBas1}  I. D. Ivanovic. ``Geometrical description of quantum
state determination'' Journal of Physics A, 14, 3241-3245 (1981).
\bibitem{unbiaBas2}W.K. Wootters, and B.D. Fields, ``Optimal state-determination by mutually
unbiased measurements'' Ann. Phys. 191, 363-381 (1989).
\bibitem{unbiaBas3} S. Bandyopadhyay, P. Boykin, V. Roychowdhury,
and F. Vatan, ``A new proof for the existence of mutually
unbiased bases'' Algorithmica 34, 512 (2002). arXiv:
quant-ph/0103162.
\bibitem{plast}A. Majtey, P. W. Lamberti, M. T. Martin and A. Plastino,
``Wootter's distance revisited: a new distinguishability
criterium'' Phys. Lett. A \textbf{32}, 413-419 (2005). arXiv:
quant-ph/0408082
\bibitem{haus} E. Lages Lima, \emph{Espa\c{c}os M{\'e}tricos},
Projeto Euclides, IMPA, ISBN: 85-244-0158-3 3Ed. Rio de Janeiro,
2003.
\bibitem{findim1} A. C. de la Torre, D. Goyeneche.
``Quantum mechanics in finite dimensional Hilbert space''
 Am. J. Phys. {\bf71}, 49-54, (2003).
\bibitem{findim2}A. C. de la Torre, H. M{\'a}rtin, D. Goyeneche.
``Quantum diffusion on a cyclic one dimensional lattice'' Phys.
Rev. E {\bf 68}, 031103-1-9, (2003).
\bibitem{wei4} J. P. Amiet, S. Weigert. ``Reconstructing a pure state of a spin s
through three Stern-Gerlach measurements'' J. Phys. A: Math. Gen.
\textbf{32}, 2777-2784 (1999).
\end{thebibliography}
\end{document}